\documentclass[article]{jss}
\usepackage{amsmath}
\usepackage{booktabs}
\usepackage{float}
\usepackage{graphicx}
\usepackage{tikz}

\usepackage{bm}

\graphicspath{ {./plots/} }
\usepackage[T1]{fontenc}

\def\bbeta{{\bm\beta}}
\def\by{{\bf y}}
\def\ones{{\bf 1}}

\author{Alex Boyd\\University of California,\\Irvine \And 
        Dennis L. Sun\\California Polytechnic\\State University}
\title{\pkg{salmon}: A Symbolic Linear Regression Package for \proglang{Python}}

\Plainauthor{Alex Boyd, Dennis L. Sun} 
\Plaintitle{salmon: A Symbolic Linear Modeling Package for Python} 
\Shorttitle{\pkg{salmon}: A Linear Modeling Package} 

\Abstract{
  One of the most attractive features of \proglang{R} is 
  its linear modeling capabilities. We describe a 
  \proglang{Python} package, \pkg{salmon}, that brings the 
  best of \proglang{R}'s linear modeling functionality 
  to \proglang{Python} in a Pythonic way---by providing 
  composable objects for specifying and fitting linear 
  models. This object-oriented design also enables other 
  features that enhance ease-of-use, such as 
  automatic visualizations and intelligent model building.
}
\Plainkeywords{linear regression, linear model, Python} 

\Address{
  Dennis Sun\\
  Department of Statistics\\
  California Polytechnic State University\\
  1 Grand Ave\\
  San Luis Obispo, CA 93407\\
  E-mail: \email{dsun09@calpoly.edu}\\
  URL: \url{http://dlsun.github.io/}
}

\def\bx{{\bf x}}

\begin{document}


\section[Introduction]{Introduction}
Linear models are ubiquitous in 
statistics, science, engineering, and machine learning. A linear 
model assumes that the expected value of a response variable $y$ is 
a linear function of explanatory variables, $x_1, ..., x_p$:
\begin{equation}
E[y|{\bf x}] = \beta_0 + \beta_1 x_1 + \beta_2 x_2 + ... + \beta_p x_p
\label{eq:linear_regression}
\end{equation}
for coefficients $\beta_0, \beta_1, ..., \beta_p$. Model 
\eqref{eq:linear_regression} is more flexible and general than 
it may first appear. For example, linear models do not necessarily have 
to be linear in the original explanatory variables. If we add higher-order 
polynomial terms to a linear model, then we can also model non-linear effects:
\begin{equation}
E[y|{\bf x}] = \beta_0 + \beta_1 x_1 + \beta_2 x_1^2 + \beta_3 x_1^3 + \beta_4 x_2.
\label{eq:polynomial_regression}
\end{equation}

How are linear models used?
First, they can be used for {\bf description}. 
For example, we may want to emphasize trends in a scatterplot by 
superimposing a best-fit line on the points. Second, they may be used 
for {\bf prediction}. Once the coefficients $\beta_0, ..., \beta_p$ 
have been estimated, the linear model can be used to predict the 
value of the response $y$ for a 
new observation where only the explanatory variables are known. 
Finally, linear regression can be used for 
{\bf inference}. The coefficients may encode information
about nature, such as the causal effect of one variable on another, in 
which case we need hypothesis tests and confidence 
intervals for model parameters.

\begin{figure}
    \centering
    \begin{tikzpicture}
    
    \draw[->] (-0.2, 0) -- (4.1, 0);
    \node[anchor=north] at (3.8, 0) {Flexibility};
    \draw[->] (0, -0.2) -- (0, 4.1);
    \node[anchor=east] at (0, 4) {Ease of Use};
    
    \node at (0.5, 3.5) {\proglang{JMP}};
    \node at (1, 3) {\proglang{Minitab}};
    
    \node at (3.5, 0.4) {\proglang{MATLAB}};
    \node at (3.5, 1.1) {\pkg{scikit-learn}};
    \node at (3.5, 1.8) {\proglang{R}};
    \node at (3.5, 1.6) {\pkg{statsmodels}};
    \node at (3.5, 2.5) {\pkg{salmon}};
    
    \end{tikzpicture}
    \caption{Tradeoffs of different solutions for fitting linear models}
    \label{fig:tradeoffs}
\end{figure}
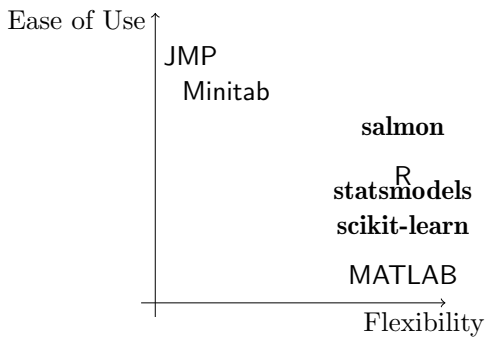

Because of these different use cases, several solutions for fitting 
linear models have emerged. Figure~\ref{fig:tradeoffs} illustrates the 
tradeoffs. At one extreme are point-and-click software packages, like \proglang{JMP}
\citep{jmp} and \proglang{Minitab} \citep{minitab}, which fit linear 
models and provide automatic visualizations. Although these 
packages have much built-in functionality, they are not 
easily extensible. For example, complicated data cleaning and wrangling often 
have to be done outside these software packages.

At the other extreme are programming languages for scientific 
computing, like \proglang{MATLAB} \citep{matlab} and \proglang{Python} \citep{python}. 
To fit a linear model, users have to manually construct 
the matrices to be passed to a least-squares solver. To obtain 
predictions from the fitted model, users must implement the 
matrix multiplications. Although these environments can be 
powerful, users have to keep track of low-level details 
that distract from the modeling. 

Libraries within these languages ease the burden somewhat. 
For example, \pkg{scikit-learn} is a \proglang{Python} library for machine learning that provides a consistent API for specifying, fitting, and 
predicting using a linear regression model. \citep{pedregosa2011scikit} However, it provides little 
help with the other two uses of linear regression, 
description and inference,
offering neither visualizations nor uncertainty estimates. 
Also, the \code{LinearRegression} model in \pkg{scikit-learn} assumes 
that the explanatory variables have already been transformed into the 
numerical matrix that will be passed into a least-squares solver, so 
users must manually transform the variables or else define 
the transformations as part of the model.

\proglang{R} occupies a medium between the two extremes. \citep{R}
On the one hand, it is a full-fledged programming language. 
On the other, it provides a high-level API for specifying and fitting 
linear regression models through formulas and the \code{lm} function. 
For example, model \eqref{eq:polynomial_regression} 
could be specified and fit in \proglang{R} as 
\begin{center}
\begin{CodeInput}
lm(y ~ poly(x1, 3) + x2, data)
\end{CodeInput}
\end{center}
Although \proglang{R} provides automatic diagnostic plots, it 
offers limited visualizations of the fitted model, in comparison 
with point-and-click software packages such as \proglang{JMP} and 
\proglang{Minitab}.

\pkg{Patsy} \citep{patsy} and \pkg{statsmodels} \citep{seabold2010statsmodels}  are \proglang{Python} libraries that 
port \proglang{R}-style modeling to \proglang{Python}. Like \proglang{R},
\pkg{statsmodels} provides some automatic diagnostic plots but not 
visualizations of the fitted model.
However, \proglang{R} formula syntax is not 
interpretable by \proglang{Python}, so formulas have to be specified as 
strings:
\begin{center}
\begin{CodeInput}
smf.ols("y ~ poly(x1, 3) + x2", data)
\end{CodeInput}
\end{center}
This creates problems when the column names do not meet the language's 
rules for variable names. For example, columns with names such as 
\code{"weight.in.kg"} or \code{"person's height"} would need to be wrapped in Patsy's ``quote'' object, \code{Q}:
\begin{center}
\begin{CodeInput}
smf.ols("y ~ Q('weight.in.kg') + Q('person\'s height')", data)
\end{CodeInput}
\end{center}
Because the \code{Q} constructor takes a string as input, we have
strings within strings---hence the need for two different kinds of 
string delimiters. If the variable name itself includes an apostrophe 
or a quotation mark, then that character needs to be escaped. An
object-oriented approach to model specification, described in this 
paper, avoids these complications, allowing any object that is a valid 
column name in a \pkg{pandas} \code{DataFrame} to be used directly 
in the model specification, without requiring a workaround for 
problematic variable names. Furthermore, having objects associated with 
each variable in a model makes it easy to specify customization 
options for each variable (e.g., the baseline level for a 
categorical variable).


In this paper, we describe \pkg{salmon}, a 
\proglang{Python} package for 
linear regression that offers an object-oriented interface 
for specifying linear models. The two main contributions of \pkg{salmon} are:
\begin{enumerate}
\item providing an 
\proglang{R}-like (but Pythonic) API for specifying models,
\item producing appropriate visualizations of the 
models, bridging the gap with point-and-click packages, like \proglang{JMP} 
and \proglang{Minitab}.
\end{enumerate}
Its philosophy and design is similar 
to other statistical packages in \proglang{Python}, such as \pkg{Symbulate} \citep{ross2019symbulate}.
Throughout, we provide comparisons of 
model specification in \pkg{salmon} with the 
similar formula syntax of \pkg{R}.

The easiest way to obtain \pkg{salmon} 
is to install it via \code{pip}:
\begin{CodeInput}
pip install salmon-lm
\end{CodeInput}
but it can also be installed 
from source at 
\url{http://github.com/ajboyd2/salmon}.


\section[Model building and fitting]{Model building and fitting}

\label{sec:model_fitting}


We introduce the design and syntax of 
\pkg{salmon} by way of a case study. We 
assume that all \pkg{salmon} objects and 
methods have already been imported into 
the global namespace, as follows:

\begin{CodeInput}
from salmon import Q, C, Log, Poly, LinearModel
\end{CodeInput}

We will use a sampled version of the 
Ames housing data set \citep{de2011ames}, which can be found within the repository for the package. The first five rows of this sampled 
data set are shown in Table~\ref{tbl:data}.

\begin{table}[H]
\begin{center}
\begin{tabular}{llrlrl}
\toprule
{} & Neighborhood &  Price(\$) & Style &  Sq. Ft. & Fire? \\
\midrule
0 &      SawyerW &         162000 &     2 Story &            1400 &         No \\
1 &      CollgCr &         195000 &     2 Story &            1660 &         No \\
2 &      Crawfor &         164000 &       Other &            1646 &        Yes \\
3 &      NridgHt &         417500 &     1 Story &            2464 &        Yes \\
4 &      SawyerW &         186800 &     1 Story &            1400 &         No \\
\bottomrule
\end{tabular}

\caption{First five rows of the Ames Housing dataset, with only the relevant columns for this paper shown.}
\label{tbl:data}
\end{center}
\end{table}

We will start with the simplest possible 
model, which assumes the sale price is a 
linear function of just the square footage:

\begin{equation}
E[\text{Price(\$)}|{\bf x}] = \beta_0 + \beta_1 (\text{Sq. Ft.}).
\label{eq:model_1}
\end{equation}

This simple linear regression model can be specified 
in \pkg{salmon} as follows:

\begin{CodeInput}
>>> x = Q("Sq. Ft.") 
>>> y = Q("Price($)")
>>> simple_model = LinearModel(x, y)
>>> print(simple_model)
Price($) ~ 1 + Sq. Ft.
\end{CodeInput}

Notice that a quantitative variable is specified by creating 
a \code{Quantitative} object, or \code{Q} for short, with the name of the 
column in the \code{DataFrame}. Alternatively, 
we could have created a generic variable using \code{V} and 
let \pkg{salmon} infer the type. 
Either way, these variable objects become the 
explanatory (\code{x}) and response (\code{y}) components of a \code{LinearModel} object.

An intercept is added by default, as evidenced 
by the constant term \code{1} in the printout. To specify 
a model without an intercept, we could either insert 
\code{- 1} into the expression 
(which mirrors the formula syntax of \proglang{R}) 
or specify \code{intercept=False}:

\begin{CodeInput}
>>> no_intercept_model = LinearModel(x - 1, y) 
>>> no_intercept_model = LinearModel(x, y, intercept=False)
>>> print(no_intercept_model)
Price($) ~ Sq. Ft.
\end{CodeInput}

To fit the \code{simple_model} above to data, we 
call the \code{.fit()} method and pass in a \pkg{pandas} 
\code{DataFrame} containing those variables:

\begin{CodeInput}
>>> simple_model.fit(data)
\end{CodeInput}

\begin{table}[H]
\begin{center}
\begin{tabular}{lrrrrrr}
\toprule
{} &  Coefficient &    SE &     t &      p &  2.5\% &  97.5\% \\
\midrule
Sq. Ft. &        118.5 & 2.068 & 57.33 &  0.000 & 114.5 &  122.6 \\
Intercept      &         3614 &  3254 & 1.111 & 0.2668 & -2766 &   9995 \\
\bottomrule
\end{tabular}

\caption{Coefficients and inferences for model~\eqref{eq:model_1},  \code{\detokenize{simple_model}}.}
\end{center}
\end{table}

Notice that \code{.fit()} returns the standard regression 
output---containing the coefficients, their standard errors, the $t$-statistic and $p$-value for testing $\beta_j = 0$, and associated 95\% confidence intervals---stored in a \code{DataFrame} for easy display and access. From the 
regression output, 
\text{Sq. Ft.} appears to have substantial explanatory power, 
but to be 
sure, we should visualize the model. A model can be 
visualized using the \code{.plot()} method; \pkg{salmon} will 
automatically choose an appropriate visualization:

\begin{CodeInput}
>>> simple_model.plot()
\end{CodeInput}

\begin{figure}[H]
\centering
\includegraphics{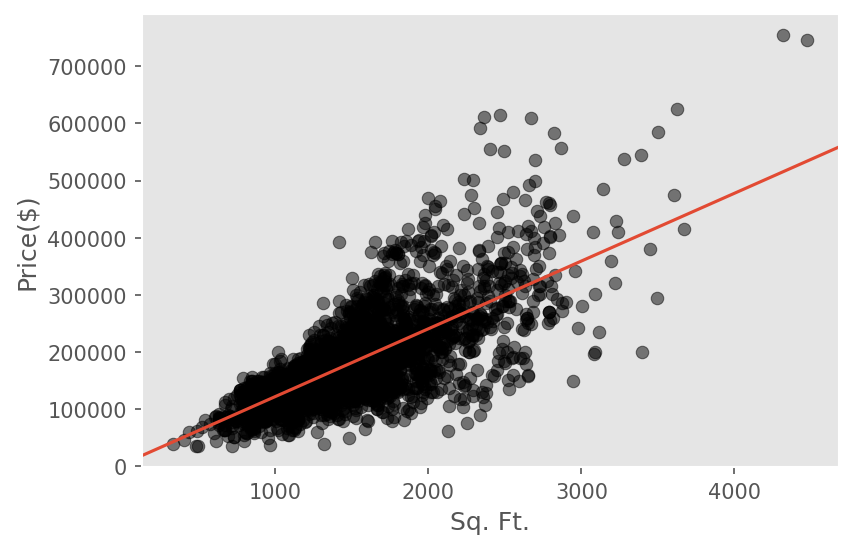}
\caption{Basic visualization of the fitted model~\eqref{eq:model_1}, \code{\detokenize{simple_model}}.}
\label{simple_plot_no_conf}
\end{figure}

Notice that the line of code above produces a scatterplot 
of the data in black, with the fitted regression 
line superimposed 
in red (Figure~\ref{simple_plot_no_conf}). To produce a similar plot from a 
model that was fit in 
\pkg{R}, \pkg{scikit-learn}, or \pkg{statsmodels}, 
we would have had to first create a grid of 
``test'' $x$ values, use the fitted model to predict the value of the 
response at each of those $x$ values, and then plot 
these predictions as a line using some plotting library.

Confidence and prediction intervals can be added to any 
model visualization, by passing the desired error rate $\alpha$ 
(so that the confidence level is $1 - \alpha$) to arguments \code{confidence_band=} or \code{prediction_band=} 
in \code{.plot()}:
\begin{CodeInput}
>>> alpha_val = 0.05
>>> simple_model.plot(confidence_band=alpha_val) 
\end{CodeInput}

\begin{figure}[H]
\centering
\includegraphics{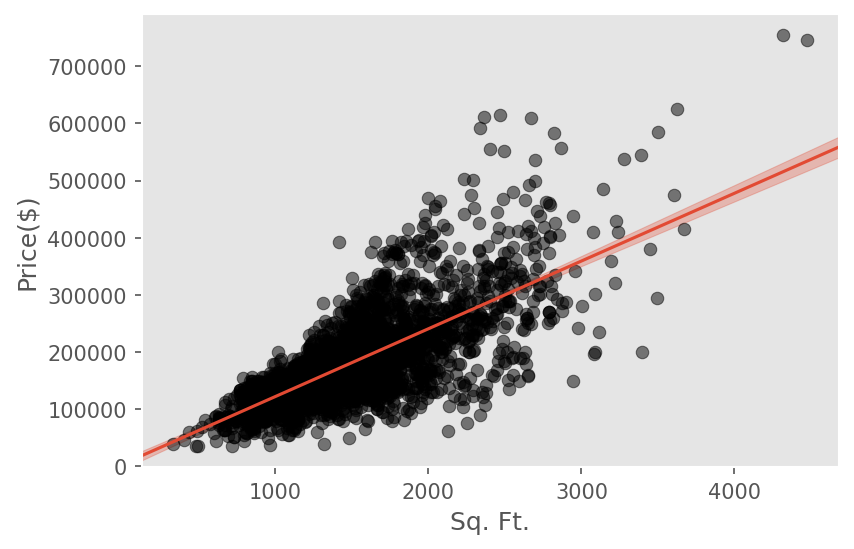}
\caption{Visualization of the fitted model~\eqref{eq:model_1}, \code{\detokenize{simple_model}}, with confidence band.}
\label{simple_plot_conf}
\end{figure}


In Figure~\ref{simple_plot_conf}, the points appear to ``fan out'' from 
the fitted model as square footage increases. This suggests 
that the assumption of constant variance (homoskedasticity) 
may be violated. One fix is to 
transform the response variable. In other words, we can 
instead fit the model:
\begin{equation}
E[\log(\text{Price(\$)})|{\bf x}] = \beta_0 + \beta_1 (\text{Sq. Ft.})
\label{eq:model_2}
\end{equation}
which can be accomplished by literally transforming the 
response:
\begin{CodeInput}
>>> simple_log_model = LinearModel(Q("Sq. Ft."), Log(Q("Price($)")))
>>> print(simple_log_model)
log(Price($)) ~ 1 + Sq. Ft.
\end{CodeInput}
Compare with \proglang{R}, where the same model would be specified as 
\begin{CodeInput}
log(`Price($)`) ~ `Sq. Ft.`
\end{CodeInput}
Note that the backticks are only necessary due to the variable names containing special characters (e.g., spaces and dollar sign). 

\pkg{salmon} supports a number of different transformations by default, such as:

\begin{itemize}
    \item Natural Logarithm: {\tt Log(X)},
    \item Logarithm of Base 10: {\tt Log10(X)},
    \item Sine: {\tt Sin(X)},
    \item Cosine: {\tt Cos(X)},
    \item Exponential: {\tt Exp(X)},
    \item Standardization: {\tt Std(X), Standardize(X)},
    \item Centering: {\tt Cen(X), Center(X)},
    \item Identity: {\tt Identity(X)}.
\end{itemize}

Now we can fit and visualize this new model:
    
\begin{CodeInput}
>>> simple_log_model.fit(data)
\end{CodeInput}

\begin{table}[H]
\begin{center}
\begin{tabular}{lrrrrrr}
\toprule
{} &  Coefficient &       SE &     t &     p &     2.5\% &    97.5\% \\
\midrule
Sq. Ft. &     5.906e-4 & 1.055e-5 & 55.96 & 0.000 & 5.699e-4 & 6.113e-4 \\
Intercept      &        11.14 &   0.0166 & 670.6 & 0.000 &    11.11 &    11.17 \\
\bottomrule
\end{tabular}

\caption{Coefficients and inferences for the fitted model~\eqref{eq:model_2},  \code{\detokenize{simple_log_model}}.}
\end{center}
\end{table}

\begin{CodeInput}
>>> simple_log_model.plot(confidence_band=alpha_val)
\end{CodeInput}

\begin{figure}[H]
\centering
\includegraphics{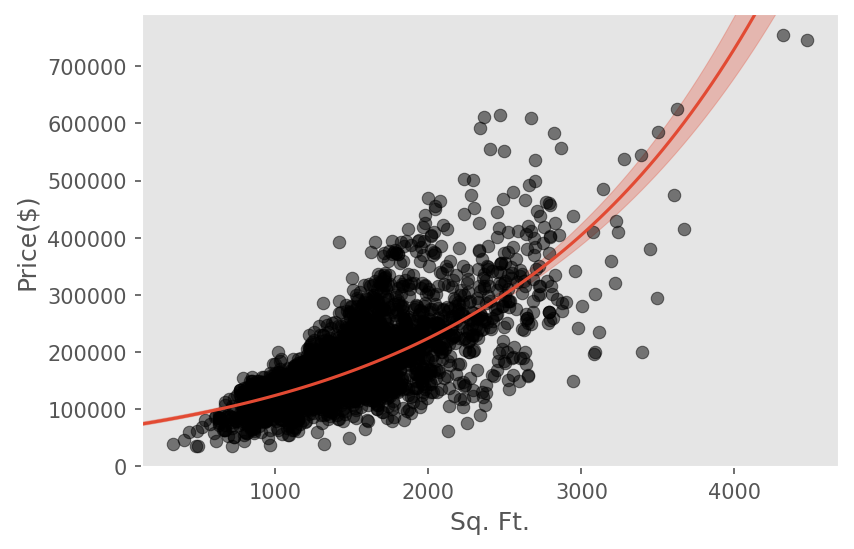}
\caption{Visualization of the fitted model~\eqref{eq:model_2}, \code{\detokenize{simple_log_model}}, with confidence band. By default, \pkg{salmon} plots the original variables, rather than the transformed variables.}
\label{simple_trans_plot_conf}
\end{figure}

By default, \pkg{salmon} plots the variables in the original 
(untransformed) space, which is usually desired. After all, 
we are interested in $\text{Price(\$)}$, not $\log(\text{Price(\$)})$. 
However, the log-space view can be  
helpful for checking whether the linear regression assumptions are met. To produce a plot in the transformed space, specify 
\code{transformed_y_space=True}.

\begin{CodeInput}
>>> simple_log_model.plot(confidence_band=alpha_val, transformed_y_space=True)
\end{CodeInput}

\begin{figure}[H]
\centering
\includegraphics{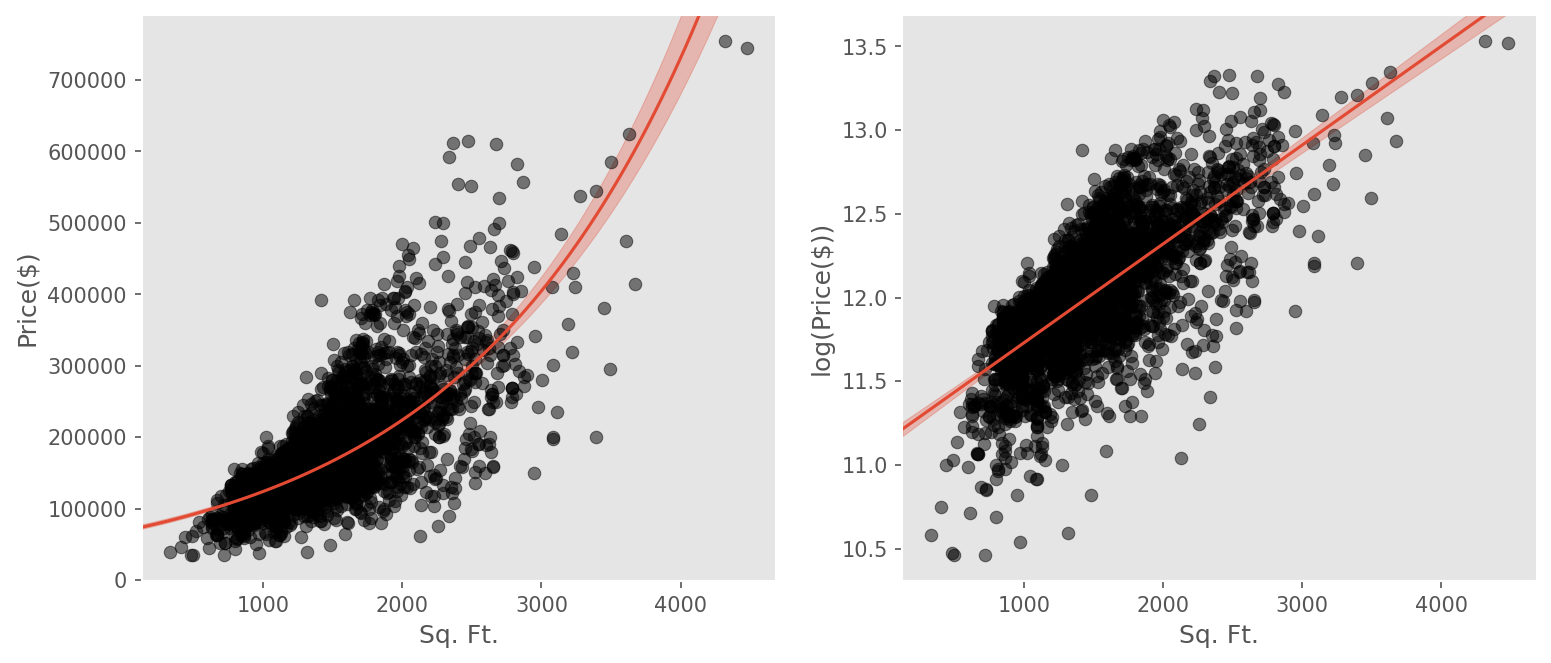}
\caption{Two visualizations of model~\eqref{eq:model_2}, \code{\detokenize{simple_log_model}}. The plot on the left shows the original response ($\text{Price(\$)}$), while the plot on the right shows the transformed response ($\log(\text{Price(\$)})$).}
\label{simple_trans_plot_conf_both}
\end{figure}


From Figure~\ref{simple_trans_plot_conf_both}, it is clear that the fanning is 
indeed reduced in log-space---as we had hoped! However, the model 
may be further improved by adding a quadratic term 
to capture the parabolic relationship. 
We create a new model for this:
\begin{equation}
E[\log(\text{Price(\$)})|{\bf x}] = \beta_0 + \beta_1 (\text{Sq. Ft.}) + \beta_2 (\text{Sq. Ft.})^2
\label{eq:model_3}
\end{equation}

A polynomial model, like \eqref{eq:model_3}, can be 
specified in \pkg{salmon} using the \code{Poly()} class, much as one uses the \code{poly()} function 
in \pkg{R}:

\begin{CodeInput}
>>> poly_model = LinearModel(
        Poly(Q("Sq. Ft."), 2), 
        Log(Q("Price($)")))
>>> print(poly_model)
Log(Price($)) ~ 1 + Sq. Ft. + (Sq. Ft.)^2
>>> poly_model.fit(data)
\end{CodeInput}


\begin{table}[H]
\begin{center}
\begin{tabular}{lrrrrrr}
\toprule
{} &  Coefficient &       SE &      t &       p &      2.5\% &     97.5\% \\
\midrule
Sq. Ft.   &     9.989e-4 & 4.603e-5 &  21.70 & 5.9e-97 &  9.087e-4 &    0.0011 \\
Sq. Ft.\textasciicircum 2 &    -1.180e-7 & 1.296e-8 & -9.107 & 1.5e-19 & -1.434e-7 & -9.261e-8 \\
Intercept        &        10.82 &   0.0386 &  280.3 &   0.000 &     10.75 &     10.90 \\
\bottomrule
\end{tabular}

\caption{Coefficients and inferences for model~\eqref{eq:model_3},  \code{\detokenize{poly_model}}.}
\end{center}
\end{table}

As can be seen, in this example \code{Poly(Q("Sq. Ft.", 2)} is equivalent to 
\code{Q("Sq. Ft.") + Q("Sq. Ft.") ** 2}.

\begin{CodeInput}
>>> poly_model.plot(confidence_band=alpha_val, transformed_y_space=True)
\end{CodeInput}

\begin{figure}[H]
\centering
\includegraphics{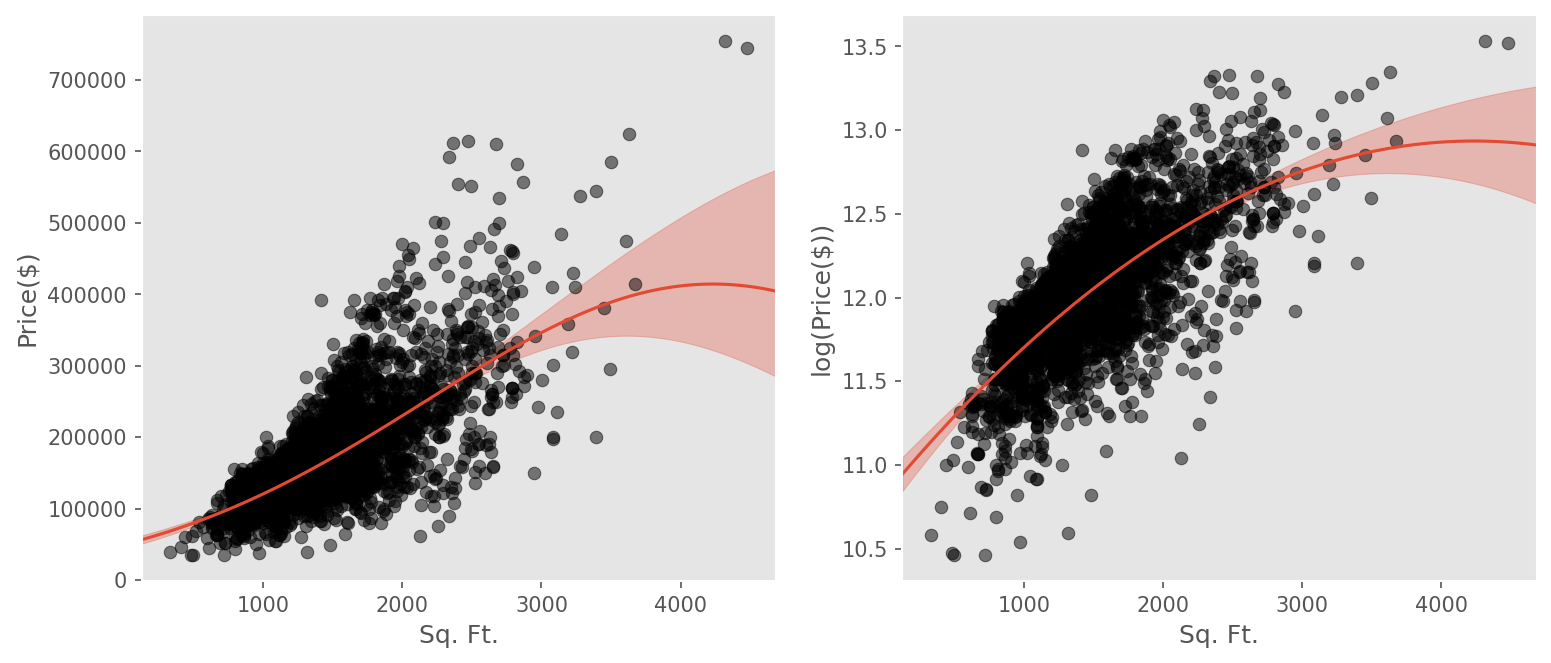}
\caption{Two visualizations of the fitted 
polynomial model~\eqref{eq:model_3},  \code{\detokenize{poly_model}}. The plot on the left shows the original response (Sale Price), while the plot on the right shows the transformed response ($\log(\text{Price(\$)})$).}
\label{poly_trans_both_conf}
\end{figure}

The $p$-value for the quadratic term $\text{(Sq. Ft.)}^2$ is 
practically zero, which suggests that the fit is substantially 
improved by adding the quadratic term.

So far, the models have only used square footage. It is 
worth investigating whether the fit can be improved by adding the two categorical variables 
in the data set---the presence of a fireplace and the overall style of house. The former is binary, with a value 
of ``Yes" indicating that the house has a fireplace. The latter can take three different values: ``1 Story'', ``2 Story'' and ``Other''. 
\begin{equation}
\begin{split}
E[\log(\text{Price(\$)})|{\bf x}] = & \beta_0 + \beta_1 I(\text{Style=``2 Story"}) + \beta_2 I(\text{Style=``Other"}) + \\
& \beta_3 I(\text{Fire?=``Yes"})
\label{eq:model_4}
\end{split}
\end{equation}
where $I(\cdot)$ denotes the indicator function.
\begin{CodeInput}
>>> simple_cat_model = LinearModel(
        C("Style") + C("Fire?"), 
        Log(Q("Price($)")))
>>> print(simple_cat_model)
Log(Price($)) ~ 1 + Style + Fire?
>>> simple_cat_model.fit(data)
\end{CodeInput}
The same model would be specified in \proglang{R} as
\begin{CodeInput}
log(`Price($)`) ~ `Style` + `Fire?`.
\end{CodeInput}
The backquotes are necessary because of the 
non-standard column names. Also, the code 
above assumes that 
\code{Style} and \code{Fire?} 
are unambiguously categorical. 
If they could be confused for 
quantitative variables, they would have to 
be explicitly cast to factors:
\begin{CodeInput}
log(`Price($)`) ~ as.factor(`Style`) + as.factor(`Fire?`)
\end{CodeInput}
This is an example where \pkg{salmon}'s insistence on explicit
variable types using \code{C} and \code{Q} saves typing in the 
long run.

\begin{table}[H]
\begin{center}
\begin{tabular}{lrrrrrr}
\toprule
{} &  Coefficient &     SE &      t &       p &    2.5\% &   97.5\% \\
\midrule
Style\{2 Story\} &       0.1058 & 0.0143 &  7.374 & 2.1e-13 &  0.0777 &  0.1339 \\
Style\{Other\}   &      -0.1500 & 0.0164 & -9.156 & 9.8e-20 & -0.1821 & -0.1179 \\
Fire?\{Yes\}      &       0.3966 & 0.0124 &  31.87 &  1e-191 &  0.3722 &  0.4210 \\
Intercept            &        11.82 & 0.0105 &   1124 &   0.000 &   11.80 &   11.84 \\
\bottomrule
\end{tabular}

\caption{Coefficients and inferences for model~\eqref{eq:model_4},  \code{\detokenize{simple_cat_model}}.}
\end{center}
\end{table}

\begin{CodeInput}
>>> simple_cat_model.plot(confidence_band=alpha_val, transformed_y_space=True)
\end{CodeInput}

\begin{figure}[H]
\centering
\includegraphics{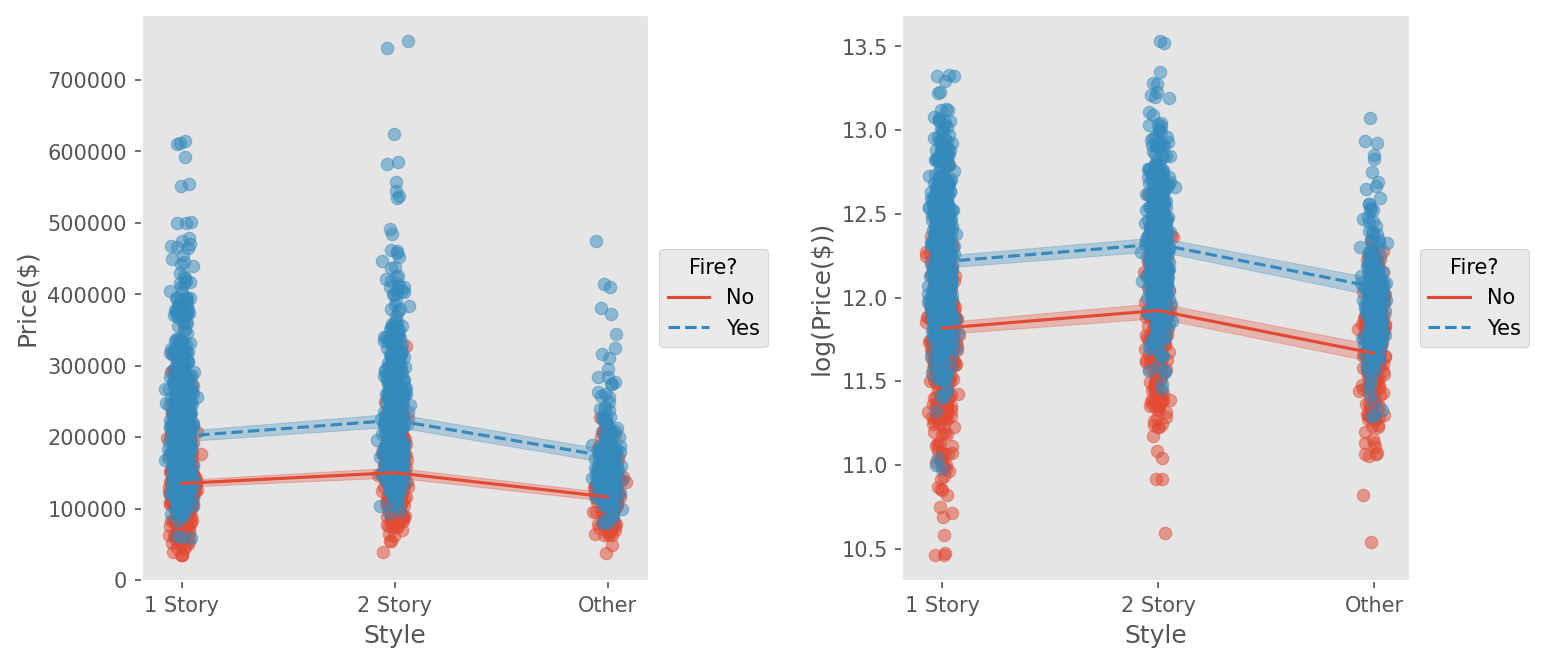}
\caption{Two visualizations of model~\eqref{eq:model_4},  \code{\detokenize{simple_cat_model}}, which contains two categorical predictors. The plot on the left shows the original response (Price(\$)), while the plot on the right shows the transformed response ($\log(\text{Price(\$)})$).}
\label{cat_conf}
\end{figure}

To specify that these variables are categorical, we create a 
\code{Categorical} object (or \code{C} for short). As 
the regression output shows, \pkg{salmon} automatically 
chooses a dummy / one-hot encoding of the levels and drops a baseline
level. This behavior can be 
customized by specifying additional arguments to 
\code{C}. For example, \pkg{salmon} 
chose ``1 Story'' to be the baseline 
level for Style. To make ``Other'' the 
baseline level, we would specify the 
variable as \code{C("Style", baseline="Other")}. Changing the 
baseline level does not affect the 
predictions from the model 
but does affect the 
interpretation of the coefficients.

Model~\eqref{eq:model_4} assumes no interaction between the factors, 
which is why the fitted lines in Figure~\ref{cat_conf} are parallel. To fit a 
model with an interaction term:

\begin{equation}
\begin{split}
E[\log(\text{Price(\$)})|{\bf x}] = & \beta_0 + \beta_1 I(\text{Style=``2 Story"}) + \\
& \beta_2 I(\text{Style=``Other"}) + \beta_3 I(\text{Fire?=``Yes"}) + \\
& \beta_4 I(\text{Style=``2 Story"}) I(\text{Fire?=``Yes"}) + \\
& \beta_5 I(\text{Style=``Other"}) I(\text{Fire?=``Yes"})
\end{split}
\label{eq:model_5}
\end{equation}

we use the \code{\&} operator to include interactions of all orders 
between the two \code{Categorical} variables:

\begin{CodeInput}
>>> house, fire = C("Style"), C("Fire?")
>>> interaction_model = LinearModel(
        house & fire,   # equivalent to: house + fire + house * fire, 
        Log(Q("Price($)")))
>>> print(interaction_model)
Log(Price($)) ~ 1 + Style + Fire? + (Style)(Fire?)
>>> interaction_model.fit(data)
\end{CodeInput}

The same model would be specified in \proglang{R} as 
\code{log(`Price($)`) ~ `Style` * `Fire?{`}}. 
Note that \proglang{R} uses \code{*} to mean 
factor crossing, rather than multiplication.

\begin{table}[H]
\begin{center}
\begin{tabular}{lrrrrrr}
\toprule
{} &  Coef. &     SE &       t &        p &    2.5\% &   97.5\% \\
\midrule
(Style\{2 Story\})(Fire?\{Yes\}) &      -0.0190 & 0.0289 & -0.6593 &   0.5097 & -0.0756 &  0.0376 \\
(Style\{Other\})(Fire?\{Yes\})   &      -0.1638 & 0.0327 &  -5.008 & 5.813e-7 & -0.2280 & -0.0997 \\
Style\{2 Story\}                    &       0.1123 & 0.0217 &   5.179 & 2.385e-7 &  0.0698 &  0.1548 \\
Style\{Other\}                      &      -0.0740 & 0.0223 &  -3.318 & 9.181e-4 & -0.1177 & -0.0303 \\
Fire?\{Yes\}                         &       0.4347 & 0.0173 &   25.13 &   3e-126 &  0.4008 &  0.4687 \\
Intercept                               &        11.80 & 0.0119 &   987.7 &    0.000 &   11.78 &   11.82 \\
\bottomrule
\end{tabular}

\caption{Coefficients and inferences for model~\eqref{eq:model_5},  \code{\detokenize{interaction_model}}.  (The labels have been abbreviated for space considerations.)}
\label{tbl:interaction_model}
\end{center}
\end{table}
  
\begin{CodeInput}
>>> interaction_model.plot(
        confidence_band=alpha_val, 
        transformed_y_space=True)
\end{CodeInput}

\begin{figure}[H]
\centering
\includegraphics{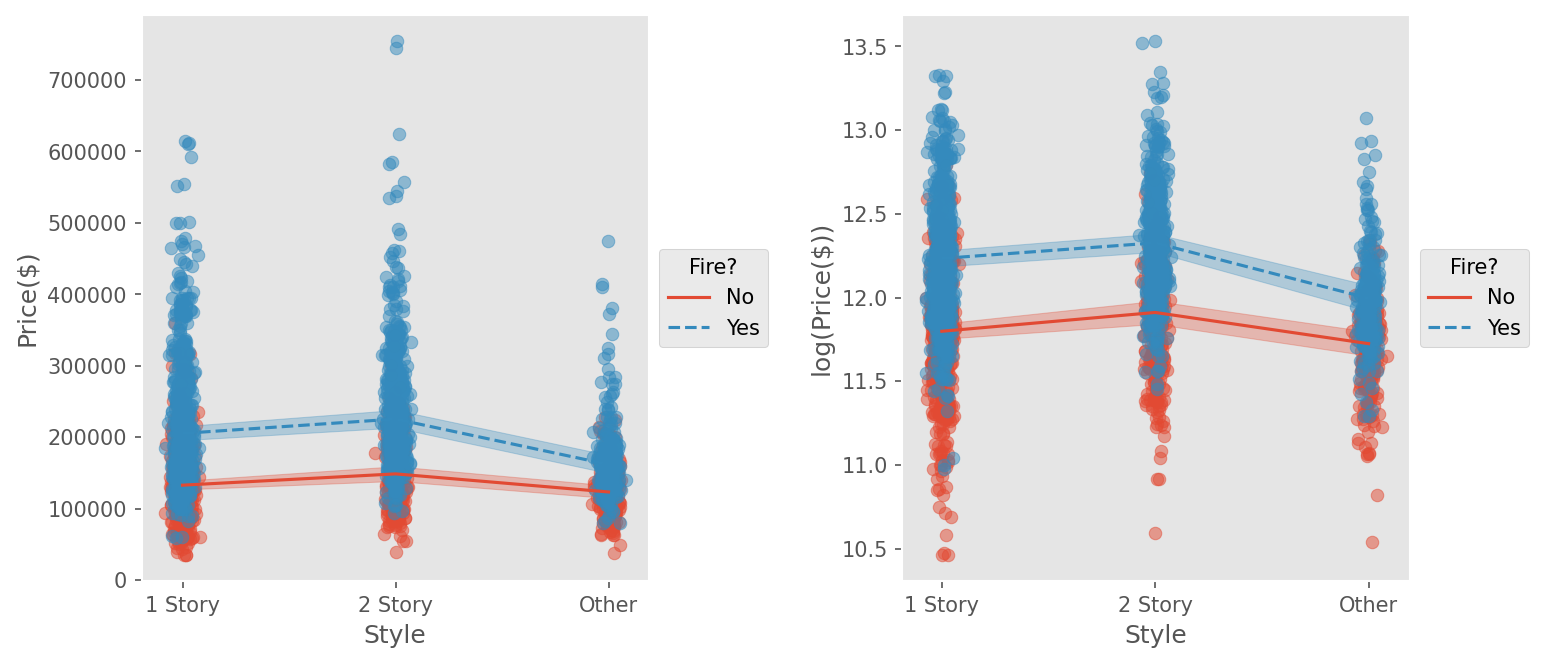}
\caption{Two visualizations of model~\eqref{eq:model_5}, \code{\detokenize{interaction_model}}, which contains two categorical predictors and their interaction. The plot on the left shows the original response (Sale Price), while the plot on the right shows the transformed response ($\log(\text{Sale Price(\$)})$).}
\label{cat_inter_conf}
\end{figure}

In Figure~\ref{cat_inter_conf}, we see that 
the visualization of the 
fitted model~\eqref{eq:model_5} is the classical interaction 
plot \citep{faraway2016linear}. Even when we allow for an interaction, the lines are still roughly 
parallel, suggesting that the interaction is weak.

Now, let's combine these categorical variables with 
the square footage into a single model. For simplicity, 
we omit the quadratic term:

\begin{equation}
\begin{split}
E[\log(\text{Price(\$)})|{\bf x}] = & \beta_0 + \beta_1 I(\text{Style=``2 Story''}) + \\
& \beta_2 I(\text{Style=``Other''}) + \beta_3 I(\text{Fire?=``Yes''}) + \\
& \beta_4 I(\text{Style=``2 Story''}) I(\text{Fire?=``Yes"}) + \\
& \beta_5 I(\text{Style=``Other''}) I(\text{Fire?=``Yes"}) + \\ 
& \beta_6 \text{(Sq. Ft.)}
\end{split}
\label{eq:model_6}
\end{equation}

Adding the quantitative variable \text{SqFt} to the model 
is straightforward:

\begin{CodeInput}
>>> house, fire, sqft = C("Style"), C("Fire?"), Q("Sq. Ft.")
>>> quant_cat_model = LinearModel(
        (house & fire) + sqft, 
        Log(Q("Price($)")))
>>> print(quant_cat_model)
Log(Price($)) ~ 1 + Sq. Ft. + Style + Fire? + (Style)(Fire?)
>>> quant_cat_model.fit(data)
\end{CodeInput}

The same model would be specified in \proglang{R} by: \\ \code{log(`Price(\$)`) ~ `Sq. Ft.` + `Style`*`Fire?{`}}.

\begin{table}[H]
\begin{center}
\begin{tabular}{lrrrr}
\toprule
{} &  Coefficient &       SE &      t &        p \\
\midrule
(Style\{2 Story\})(Fire?\{Yes\}) &      -0.0389 &   0.0215 & -1.807 &   0.0709\\
(Style\{Other\})(Fire?\{Yes\})   &      -0.1478 &   0.0244 & -6.057 & 1.564e-9 \\
Style\{2 Story\}                    &      -0.1524 &   0.0171 & -8.926 &  7.7e-19 \\
Style\{Other\}                      &      -0.1535 &   0.0167 & -9.184 &  7.7e-20 \\
Sq. Ft.                          &     5.732e-4 & 1.188e-5 &  48.24 &    0.000 \\
Fire?\{Yes\}                         &       0.2160 &   0.0137 &  15.79 &  5.7e-54 \\
Intercept                               &        11.15 &   0.0161 &  691.6 &    0.000 \\
\bottomrule
\end{tabular}

\caption{Coefficients and inferences for model~\eqref{eq:model_6},  \code{\detokenize{quant_cat_model}}. (The confidence intervals are not shown for space considerations.)}
\end{center}
\end{table}

\begin{CodeInput}
>>> quant_cat_model.plot(confidence_band=alpha_val)
\end{CodeInput}

\begin{figure}[H]
\centering
\includegraphics{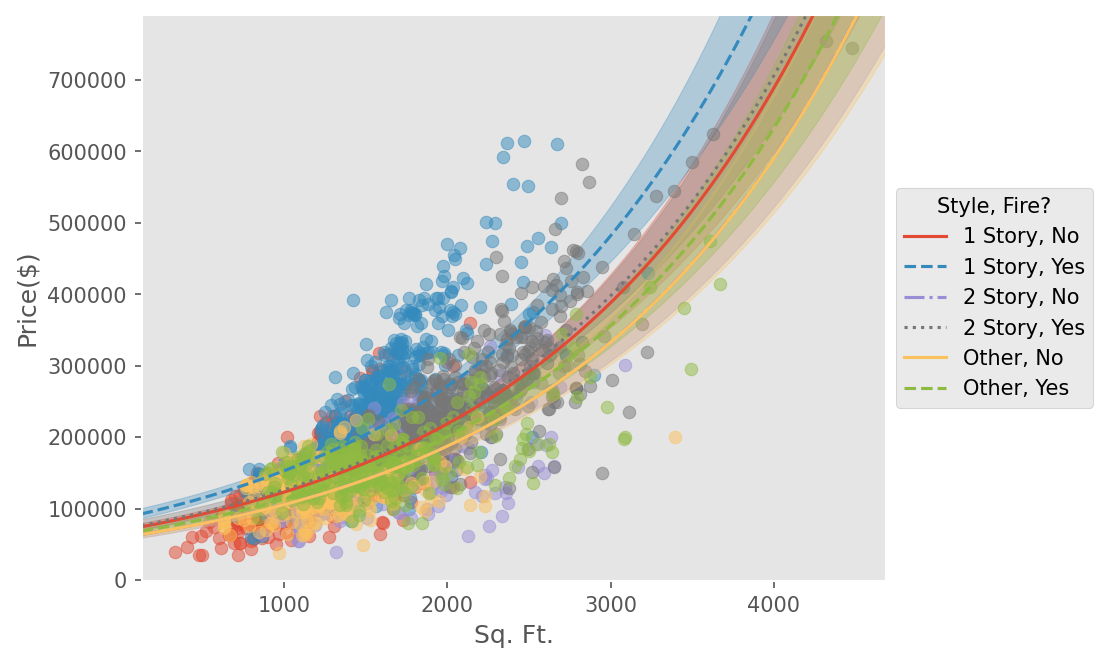}
\caption{Visualization of the fitted model~\eqref{eq:model_6}, \code{\detokenize{quant_cat_model}}. The quantitative variable is on the $x$-axis, while each unique combination of the categorical variables is represented by a different line and color.}
\label{cat_quant_trans_both_conf}
\end{figure}

However, model~\eqref{eq:model_6} 
assumed 
no interaction between the 
quantitative variable and the categorical variables. As one 
final step, we consider interacting the square footage with 
the categorical variables, which is equivalent to fitting 
a separate model for each possible combination of the 
categorical variables:

\begin{equation}
\begin{split}
E[\log(\text{Price(\$)})|{\bf x}] = & \beta_0 + \beta_1(\text{Sq. Ft.}) \beta_2 I(\text{Style=``2 Story"}) + \\
& \beta_3 I(\text{Style=``Other"}) + \beta_4 I(\text{Fire?=``Yes"}) + \\
& \beta_5 I(\text{Style=``2 Story"}) I(\text{Fire?=``Yes"}) + \\
& \beta_6 I(\text{Style=``Other"}) I(\text{Fire?=``Yes"}) + \\
& \beta_7 (\text{Sq. Ft.})I(\text{Style=``2 Story"}) + \\
& \beta_8 (\text{Sq. Ft.})I(\text{Style=``Other"}) + \\
& \beta_9 (\text{Sq. Ft.})I(\text{Fire?=``Yes"}) + \\
& \beta_{10} (\text{Sq. Ft.})I(\text{Style=``2 Story"}) I(\text{Fire?=``Yes"}) + \\ & \beta_{11}(\text{Sq. Ft.}) I(\text{Style=``Other"}) I(\text{Fire?=``Yes"})
\label{eq:model_full}
\end{split}
\end{equation}

The sheer number of interaction terms would make this model difficult to specify, 
but because the variables are represented as composable objects in \pkg{salmon}, they can be stored in variables and reused, as illustrated below.

\begin{CodeInput}
>>> house, fire, sqft = C("Style"), C("Fire?"), Q("Sq. Ft.")
>>> full_model = LinearModel(
        house & fire & sqft, 
        Log(Q("Price($)")))
>>> print(full_model)
Log(Price($)) ~ 1 + Sq. Ft. + Style + Fire? +
    (Style)(Fire?) + (Sq. Ft.)(Style) + 
    (Sq. Ft.)(Fire?) + (Sq. Ft.)(Style)(Fire?)
>>> full_model.fit(data)
\end{CodeInput}

\begin{table}[H]
\begin{center}
\begin{tabular}{lrrrr}
\toprule
{} &  Coefficient &       SE &       t &        p \\
\midrule
(Sq. Ft.)(Style\{2 Story\})             &    -2.852e-4 & 4.556e-5 &  -6.260 &  4.4e-10 \\
(Sq. Ft.)(Style\{Other\})               &    -4.985e-4 & 4.912e-5 &  -10.15 &  8.4e-24 \\
(Style\{2 Story\})(Fire?\{Yes\})            &      -0.0766 &   0.0886 & -0.8640 &   0.3877\\
(Style\{Other\})(Fire?\{Yes\})              &      -0.0236 &   0.0872 & -0.2713 &   0.7862 \\
Style\{2 Story\}                               &       0.2222 &   0.0666 &   3.338 & 8.550e-4 \\
Style\{Other\}                                 &       0.4554 &   0.0621 &   7.331 &  2.9e-13 \\
(Sq. Ft.)(Style\{2 Story\})(Fire?{Yes}) &     6.001e-5 & 5.605e-5 &   1.071 &   0.2844 \\
(Sq. Ft.)(Style\{Other\})(Fire?{Yes}) &     3.096e-5 & 6.166e-5 &  0.5021 &   0.6156 \\
Fire?\{Yes\}                                    &       0.0592 &   0.0513 &   1.155 &   0.2482 \\
Sq. Ft.                                     &     7.457e-4 & 2.922e-5 &   25.52 &   8e-130\\
(Sq. Ft.)(Fire?\{Yes\})                  &     6.014e-5 & 3.821e-5 &   1.574 &   0.1156\\
Intercept                                          &        10.96 &   0.0341 &   321.1 &    0.000 \\
\bottomrule
\end{tabular}

\caption{Coefficients and inferences for model~\eqref{eq:model_full}, \code{\detokenize{full_model}}. (The confidence intervals are not shown for space considerations.)}
\end{center}
\end{table}

\begin{CodeInput}
>>> full_model.plot(confidence_band=alpha_val)
\end{CodeInput}

\begin{figure}[H]
\centering
\includegraphics{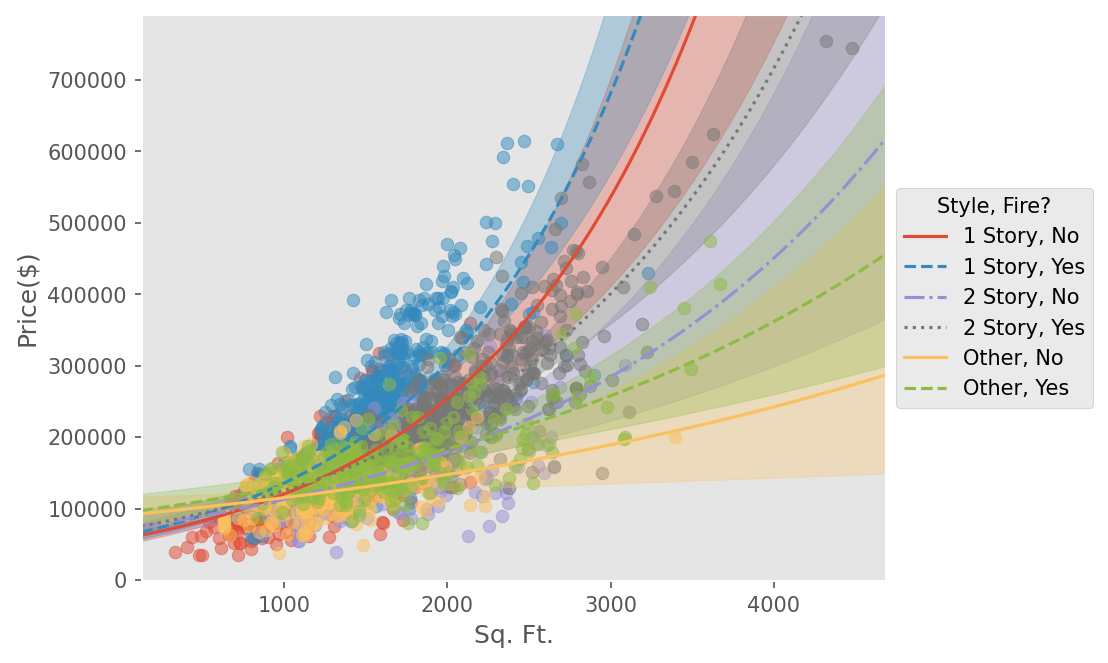}
\caption{Visualization of the fitted model~\eqref{eq:model_full}, \code{\detokenize{full_model}}, which includes an interaction between the quantitative and categorical variables.}
\label{cat_quant_inter_trans_both_conf}
\end{figure}


\proglang{R} supports a similarly concise 
syntax for the same model
\begin{CodeInput}
log(`Price($)`) ~ `Style` * `Fire?` * `Sq. Ft.`
\end{CodeInput}
but at the cost of making \code{*} mean something 
other than ``multiplication.''


To use a regression model for prediction, we can 
call the \code{.predict()} method on new data. The new 
data must be a \pkg{pandas} \code{DataFrame} whose column 
names match the variable names in the model specification. 
In most cases, the new data will be in the same format as 
the data used to fit the model, so this requirement will 
automatically be satisfied. 
In our example below, the new \code{DataFrame}, \code{new_data}, 
is simply the same as \code{data}. 

\begin{CodeInput}
>>> full_model.predict(new_data)
\end{CodeInput}

\begin{table}[H]
\begin{center}
\begin{tabular}{lr}
\toprule
{} &  Predicted log(Price(\$)) \\
\midrule
0 &                         11.82 \\
1 &                         11.94 \\
2 &                         12.00 \\
3 &                         13.00 \\
4 &                         12.00 \\
\bottomrule
\end{tabular}

\caption{First five predicted values from the \code{\detokenize{full_model}} using the \code{\detokenize{DataFrame}}: \code{\detokenize{new_data}}}
\end{center}
\end{table}

Accompanying confidence or prediction intervals can 
be obtained by passing the desired error rate $\alpha$ to 
\code{confidence_interval=} or 
\code{prediction_interval=}:

\begin{CodeInput}
>>> full_model.predict(new_data, prediction_interval=alpha_val) 
\end{CodeInput}

\begin{table}[H]
\begin{center}
\begin{tabular}{lrrr}
\toprule
{} &  Predicted log(Price(\$)) &  2.5\% &  97.5\% \\
\midrule
0 &                         11.82 & 11.36 &  12.29 \\
1 &                         11.94 & 11.48 &  12.41 \\
2 &                         12.00 & 11.54 &  12.47 \\
3 &                         13.00 & 12.53 &  13.47 \\
4 &                         12.00 & 11.53 &  12.46 \\
\bottomrule
\end{tabular}

\caption{First five predicted values with 95\% prediction intervals from the \code{\detokenize{full_model}} using the \code{\detokenize{DataFrame}}: \code{\detokenize{new_data}}.}
\end{center}
\end{table}


Occasionally, it is useful to add variables to 
a regression model without estimating their 
coefficients. In \pkg{salmon}, this is achieved 
by simply shifting the response variable.
\begin{CodeInput}
>>> offset = 3 * Q("Sq. Ft.")
>>> offset_model = LinearModel(C("Fire?"), Q("Price($)") - offset)
>>> print(offset_model)
Price($) - 3*(Sq. Ft.) ~ 1 + Fire?
\end{CodeInput}
Compare this with \pkg{R}, where an 
offset is specified using the \code{offset()} 
command:
\begin{CodeInput}
SalePrice ~ Fire + offset(3 * SqFt)
\end{CodeInput}
Anecdotally, we have evidence that including an offset term 
by shifting the response is more intuitive.
One of our colleagues, who uses \pkg{R} 
but was unfamiliar with the \code{offset()} function,
attempted to implement offsets by 
shifting the response, which does not work in \pkg{R}. 
This motivated our decision to support 
shifting the response variable in \pkg{salmon}.

\section[Model diagnostics]{Model diagnostics}


The previous section explained how to build regression models; this 
section focuses on how to evaluate them. There are both visual and 
analytical diagnostics; we only attempt to highlight a 
few examples of each. For a full description of the API, 
please refer to the online documentation.



In the following examples, we will check whether the 
assumptions are satisfied for model~\eqref{eq:model_full}, \code{full_model}, as well as compare 
it to model~\eqref{eq:model_6}, \code{quant_cat_model}, which is a nested model. 
Both models 
were defined in 
Section~\ref{sec:model_fitting}.



\subsubsection{Visual Diagnostics}

In equation \eqref{eq:linear_regression}, we only made assumptions about the conditional expectation of the 
response $y$, given the explanatory variables $\bx{}$. However, statistical inferences for linear regression 
require assumptions about the entire conditional distribution, not just the expectation. 
Perhaps the easiest way to specify this conditional distribution is to write 
\begin{equation}
    y = \beta_0 + \beta_1 x_1 + \beta_2 x_2 + ... + \beta_p x_p + \epsilon,
\end{equation}
where the error term $\epsilon$ is independent of $\bx{}$ and assumed to follow a 
$\text{Normal}(0, \sigma^2)$ distribution. One can easily verify that this model 
is a special case of 
\eqref{eq:linear_regression}.

The assumption that the errors are independent and normally distributed with constant variance is 
usually assessed by inspecting the residuals. In \pkg{salmon}, the residual diagnostic plots 
can be generated directly from the fitted model:

\begin{CodeInput}
>>> full_model.plot_residual_diagnostics()
\end{CodeInput}

\begin{figure}[H]
\centering
\includegraphics{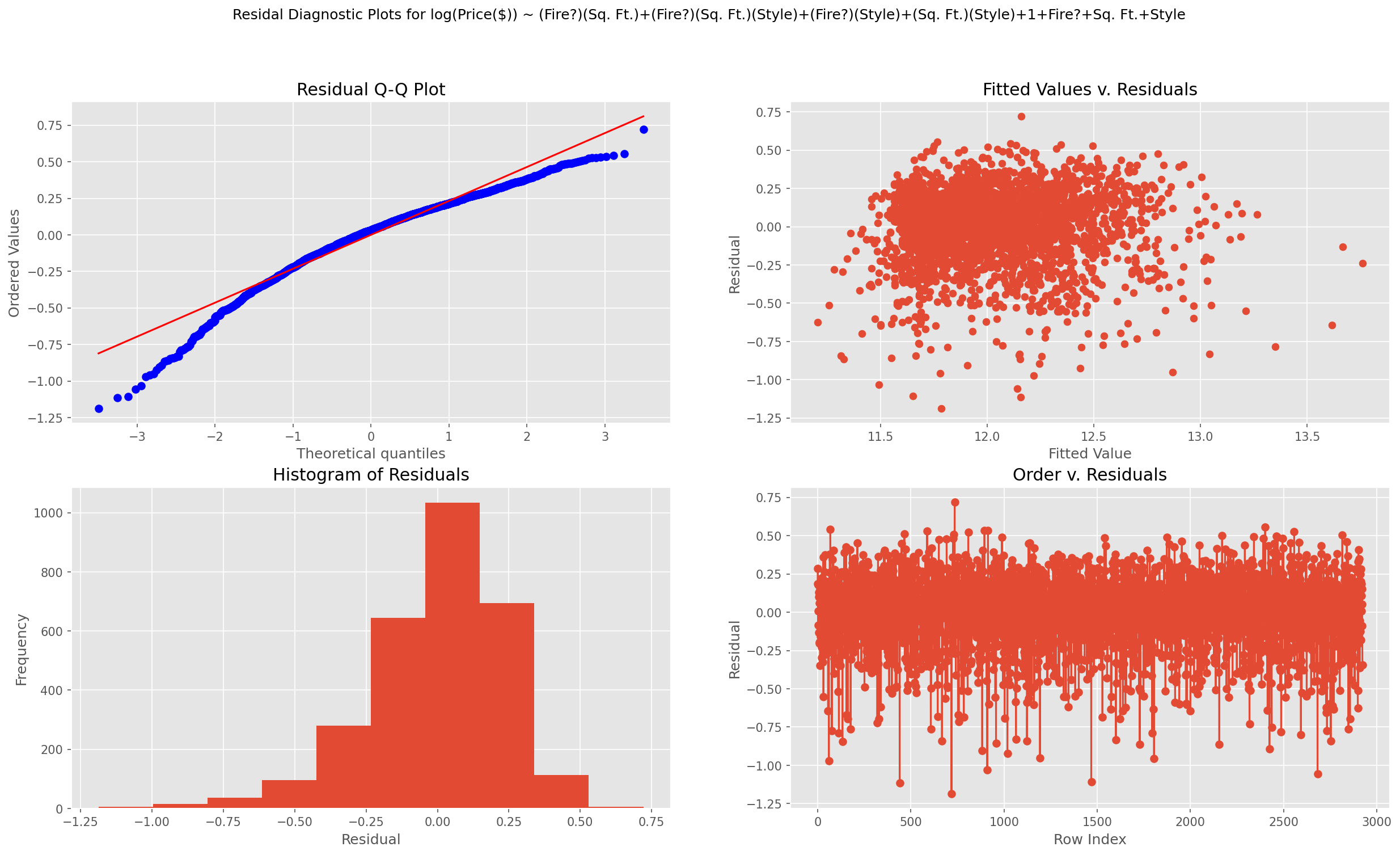}
\caption{Residual diagnostic plots for the fitted regression model.}
\label{fig:residual_diagnostics}
\end{figure}

Four plots of the residuals are produced, as shown in Figure~\ref{fig:residual_diagnostics}: 
a normal Q-Q plot, a histogram, a scatterplot of the fitted values versus the residuals, and 
a line plot of the residuals versus their order in the data set. The residuals 
appear to be skewed to the left, but 
otherwise there do not appear to be 
obvious violations of the 
assumptions. The line plot of the 
residuals versus their order is not as useful for this data set, since the row ordering 
is arbitrary, but it could potentially reveal violations of independence in data sets 
where row order is significant, such as time series data.

To further investigate assumptions, we can produce residual plots and partial regression plots by calling the methods \code{model.residual_plots()} and \code{model.partial_plots()} respectively. These will each produce one plot per explanatory variable.

\subsubsection{Analytical Diagnostics}

Visual diagnostics are in the eye of the beholder, so analytical 
diagnostics are equally important. One common, if not always the most useful,
diagnostic of a model's fit is the $R^2$ value. 

\begin{CodeInput}
>>> full_model.r_squared()
0.652748378553996
\end{CodeInput}

One problem with $R^2$ is that it is monotonically increasing in the number of 
variables in the model. A better diagnostic is Adjusted-$R^2$, which accounts for the 
number of variables. To calculate Adjusted-$R^2$, we specify the argument \code{adjusted=True}. Besides $R^2$, AIC and BIC are also 
available.


A more formal way of evaluating a model is to 
test it against another 
model. The \code{anova} command, 
when called on a single model, returns the results of 
the omnibus $F$-test, as well as the partial $F$-test for the 
each individual variable in the model.

\begin{CodeInput}
>>> from salmon import anova
>>> anova(full_model)
\end{CodeInput}

\begin{table}[H]
\begin{center}
\begin{tabular}{lrllll}
\toprule
{} &    DF &             SS Err. &             SS Reg. &                   F &       p \\
\midrule
Global Test                                 &    11 &   309.578 &   309.578 &  500.765 &   0.000 \\
- (Style)(Sq. Ft.)             &     2 &  169.819 &   303.415 &   54.828 & 4.2e-24 \\
- (Fire?)(Style)                 &     2 &   163.699 &   309.536 &  0.373 &  0.6883 \\
- Style                               &     2 &  166.776 &   306.458 &  27.755 & 1.1e-12 \\
- (Fire?)(Style)(Sq. Ft.) &     2 &  163.721 &  309.513 &  0.574 &  0.5630 \\
- Fire?                                &     1 &  163.731 &    309.503 &   1.333 &  0.2482 \\
- Sq. Ft.                            &     1 &   200.263 &  272.971 &   651.347 &  8e-130 \\
- (Fire?)(Sq. Ft.)              &     1 &  163.796 &  309.438 &   2.477 &  0.1156 \\
Error                                       &  2912 &                     &                     &                     &         \\
\bottomrule
\end{tabular}

\caption{Global $F$-test results for \code{\detokenize{full_model}}.}
\label{tbl:anova}
\end{center}
\end{table}

The inclusion of interactions between the quantitative variable (square footage) and the categorical variables (fireplace and house style) adds complexity to the model that may not be supported by the data. To test this, we can perform a
partial $F$-test comparing the full model to a 
reduced model without these interaction terms using 
the \code{anova} command again, except passing 
in two fitted models. This mirrors the behavior of the 
\code{anova()} function in \proglang{R}.



\begin{CodeInput}
>>> anova(full_model, quant_cat_model)
\end{CodeInput}

\begin{table}[H]
\begin{center}
\begin{tabular}{lrllll}
\toprule
{} &    DF &             SS Err. &             SS Reg. &                  F &       p \\
\midrule
Full Model      &    11 &  163.657 &   309.578 &                    &         \\
- Reduced Model &     5 &   179.015 &  294.219 &  54.655 & 2.0e-54 \\
Error           &  2912 &                     &                     &                    &         \\
\bottomrule
\end{tabular}

\caption{Partial $F$-test results from comparing  \code{\detokenize{full_model}} to  \code{\detokenize{quant_cat_model}}.}
\end{center}
\end{table}

The results of the partial $F$-test confirm that interactions significantly 
improve the fit of the model to the data.


\section{Automatic model building}

The \pkg{salmon} package also provides functions to 
automate the model building process. In the following examples, we use \pkg{salmon} to automatically select the ``best'' model, according to a metric of our choosing, from among the following variables: $\text{Sq. Ft.}-\text{E}[\text{Sq. Ft.}]$, $(\text{Sq. Ft.}-\text{E}[\text{Sq. Ft.}])^2$, $I(\text{Style=``2 Story"})$, $I(\text{Style=``Other"})$, $I(\text{Fire?=``Yes"})$, plus interactions between all categorical and quantitative variables. The response variable in this scenario is $\log(\text{Price(\$)})$.

To utilize the model building features of \pkg{salmon}, we 
first specify a model that contains all of the variables 
under consideration:

\begin{CodeInput}
>>> quant_vars = Poly(Cen(Q("Sq. Ft.")), 2)
>>> all_terms_model = LinearModel(
    (1 + C("Style")) * (1 + C("Fire?")) * (1 + quant_vars),
    Log(Q("Price($)")))
>>> print(all_terms_model)
log(Sale Price($)) ~ 1 + (Sq. Ft.-E(Sq. Ft.)) + 
    (Sq. Ft.-E(Sq. Ft.))^2 + Fire? + 
    (Fire?)(Sq. Ft.-E(Sq. Ft.)) + 
    (Fire?)((Sq. Ft.-E(Sq. Ft.))^2) + 
    Style + (Style)(Sq. Ft.-E(Sq. Ft.)) + 
    (Style)((Sq. Ft.-E(Sq. Ft.))^2) + 
    (Fire?)(Style) + 
    (Fire?)(Style)(Sq. Ft.-E(Sq. Ft.)) +
    (Fire?)(Style)((Sq. Ft.-E(Sq. Ft.))^2)
\end{CodeInput}

Stepwise selection methods are supported through the 
\code{stepwise} function. Forward stepwise is specified 
by \code{forward=True}, while backward stepwise is 
specified by \code{forward=False}. Additionally, we 
specify what metric to optimize (e.g., AIC, BIC, 
$R^2$) using the \code{method=} parameter. For example, 
if we use forward stepwise to optimize for BIC, we obtain 
the following ``best'' model:

\begin{CodeInput}
>>> from salmon import stepwise
>>> results = stepwise(
        full_model=all_terms_model,
        metric_name="BIC",
        forward=True,
        data=data)
>>> print(results["metric"])
BIC | -168.69617383126013
>>> print(results["best_model"])
log(Price($)) ~ 1  + 
    (Fire?)(Sq. Ft.-E(Sq. Ft.)) + 
    (Style)(Sq. Ft.-E(Sq. Ft.)) + 
    Fire? + Style + (Sq. Ft.-E(Sq. Ft.)) + 
    (Sq. Ft.-E(Sq. Ft.))^2 + 
    (Fire?)(Style)(Sq. Ft.-E(Sq. Ft.)^2)
\end{CodeInput}

The procedure produced a model associated with the lowest 
BIC score, but at the cost of violating what some would consider best modeling practices: namely, the inclusion of interactions and higher-order terms 
without main effects and lower-order 
terms \citep{cox1984interaction}. For example, 
\code{(Fire?)(Style)(Sq. Ft.-E(Sq. Ft.))\^{}2)} is included, 
but not \code{(Fire?)(Style)} or \code{(Fire?)(Style)(Sq. Ft.-E(Sq. Ft.))}.

Due the object-oriented nature of \pkg{salmon}, the package 
can keep track of inter-variable relationships. 
To employ this feature, we simply specify the parameter 
\code{naive=False} when calling the \code{stepwise} function:

\begin{CodeInput}
>>> results = stepwise(
        full_model=all_terms_model,
        metric_name="BIC",
        forward=True,
        naive=False,
        data=data)
>>> print(results["metric"])
BIC | -187.50648709382568
>>> print(results["best_model"])
log(Price($)) ~ 1 + (Fire?)(Sq. Ft.-E(Sq. Ft.)) + 
    (Fire?)(Sq. Ft.-E(Sq. Ft.)^2) +
    (Style)(Sq. Ft.-E(Sq. Ft.)) +
    (Style)(Sq. Ft.-E(Sq. Ft.)^2) + 
    Fireplace? + Style + (Sq. Ft.-E(Sq. Ft.)) + 
    (Sq. Ft.-E(Sq. Ft.)^2)
\end{CodeInput}

Now, the selected model is guaranteed 
to not violate standard model 
building practices. For example, 
if an interaction term is included, the 
main effects will be also.

\section{Integration with Python ecosystem}

The \pkg{salmon} package is deliberately built on top of 
\pkg{pandas}, in order to leverage the power of its 
\code{DataFrame} for organizing heterogeneous data. \citep{mckinney-proc-scipy-2010} It expects \pkg{pandas} 
\code{DataFrame}s as inputs, and it outputs results as 
\pkg{pandas} \code{DataFrame}s. However, 
\pkg{salmon} is also integrated with other packages 
within the \proglang{Python} data science ecosystem.

First, the \code{LinearModel} class in \pkg{salmon} implements \code{.fit()} and \code{.predict()} methods, so any \code{LinearModel} can be used 
as \pkg{scikit-learn} estimators. This means that all of \pkg{scikit-learn}'s 
routines for model selection and evaluation, 
such as cross validation, will work with models that are created in \pkg{salmon}. \citep{sklearn_api}

Second, \pkg{salmon} plots are produced using 
\pkg{matplotlib}. These plots can be further edited and 
customized in \pkg{matplotlib} because every 
plotting command in \pkg{salmon} returns 
a \pkg{matplotlib} \code{Figure}. \citep{Hunter:2007}




\section[Timing]{Timing and numerical comparisons}

We conducted timing comparisons between \pkg{salmon}, 
\pkg{statsmodels}, and \proglang{R}. The results are shown in 
Table~\ref{tab:timing}. The \pkg{salmon} package is consistently faster than 
\pkg{statsmodels} in all scenarios. It is also faster than \proglang{R} on 
large examples (with $p=1000$ predictors), although it is still 
slower than \proglang{R} on small examples, perhaps because formulas are native in 
\proglang{R}. The speed of \pkg{salmon} on large examples is likely due to 
the careful handling of the linear algebra, some details of which are 
described in Appendix~\ref{sec:model_fitting}.

\begin{table}[h]
    \centering
    \begin{tabular}{llrrr}
\toprule
Model & Task & \pkg{salmon} & \pkg{statsmodels} & \proglang{R} (\code{lm}) \\
\midrule
Simple Linear Regression & fit & 2.759 & 8.701 & 1.505 \\
\ (1 quantitative predictor) & predict & 0.607 & 2.267 & 0.350 \\
\midrule
Multiple Regression & fit & 3.637 & 15.590 & 5.098 \\
\ (4 quant. + 1 cat. predictors) & predict & 1.421 & 6.536 & 1.020\\
\midrule
 Multiple Regression & fit & 7.810 & 11.576 & 2.344 \\
\ (2 quant. + 1 cat. + interactions) & predict & 2.671 & 3.994 & 1.171 \\
\midrule
Multiple Regression & fit & 2198.1 & 7491.4 & 5280.5\\
\ (1000 quant.) & predict & 215.73 & 22.80 & 191.41 \\
\bottomrule
    \end{tabular}
    \caption{Timing comparisons of \pkg{salmon} and 
    competitors on four models. For each model, we time 
    two tasks: 1) fitting the model (including inferential statistics) 
    and 2) using the fitted model to make predictions. 
    All reported times are in milliseconds.}
    \label{tab:timing}
\end{table}

We also compare the numerical stability of \pkg{salmon}, \pkg{statsmodels}, and \proglang{R} 
on a synthetic example. 
First, we generate a grid of $n=1000$ evenly-spaced values in the range $[10^8 - 1, 10^8 + 1]$ 
for the explanatory variable $x$. Then, we generate the response variable $y$ according to 
the relation 
\begin{equation} 
y_i = 2 + 2 x_i + \epsilon_i, 
\label{eq:toy}
\end{equation}
where $\epsilon_1, ..., \epsilon_n$ are i.i.d. $\text{Normal}(0, 1)$. 
Equation~\eqref{eq:toy} describes a 
simple linear regression model with $\beta_0 = \beta_1 = 2$. The large values of $x$ make it 
nearly collinear with the intercept. 

The estimated parameters ($\hat\beta_0$ and $\hat\beta_1$) from the three 
packages are shown in Table~\ref{tab:numerical}. 
\proglang{R} fails to estimate a coefficient at all. 
The behavior of \pkg{statsmodels} is more troubling; 
it throws a warning, but proceeds to 
report incorrect values with a high degree of precision. 
Only \pkg{salmon} returns the correct values\footnote{Although the intercept is suspiciously 
    large, this is because the standard error of the intercept is very large. 
    Recall that the standard deviation of the 
    intercept is $\sigma\sqrt{\frac{1}{n} + \frac{\bar x^2}{\sum_i (x_i - \bar x)^2}} \approx 
    5471751$ for the parameters of our simulation.}, due to 
careful handling of numerical linear algebra (see Appendix~\ref{sec:model_fitting} 
for more details).

\begin{table}
    \centering
    \begin{tabular}{lccc}
\hline
 & \pkg{salmon} & \pkg{statsmodels} & \proglang{R} (\code{lm}) \\
\hline
Coefficient & 2.095 & 2.000 & NA \\
 & (0.1059) & ($3 \times 10^{-10}$) & (NA) \\
Intercept & $3046697$ & $0.00000002$ & $200000002$ \\
 & (5372320) & ($3 \times 10^{-18}$) & (0.0475) \\
\hline
    \end{tabular}
    \caption{Estimated parameters 
    (with standard errors in parentheses)
    from \pkg{salmon} and competitors
    on synthetic data simulated  
    according to \eqref{eq:toy}. Only \pkg{salmon} recovers 
    the correct coefficients.
    \proglang{R} fails to produce a coefficient at all, while 
    \pkg{statsmodels} returns the incorrect coefficients with a 
    high degree of confidence.
    }
    \label{tab:numerical}
\end{table}

The Jupyter notebooks to reproduce these experiments can be found on Github at 
\url{https://github.com/ajboyd2/salmon/tree/master/paper_outputs}.

\section[Conclusion]{Conclusion}

The \pkg{salmon} package allows for convenient and intuitive model building, using, and diagnosing for linear regression within \proglang{Python}. As our examples have demonstrated, the consistent interface allows for an linear regression experience in \proglang{Python} that reproduces---and in some cases, enhances---the experience in \proglang{R}.

Future work in \pkg{salmon} will pertain to extending the package's capabilities by implementing other types of models beyond linear regression, such as generalized linear models, as well as providing additional tests that are supported in \proglang{R}, such as linear contrasts. 

\section*{Acknowledgments} 

We are indebted to Ben Mangelsdorf, who contributed code to the \pkg{salmon} 
package and helped with the preparation of this paper. We are also grateful 
to Fernando Perez, Nicholas Russo, Jonathan Taylor, 
and John Walker for their feedback, as well as to the Bill and Linda Frost 
Fund for support.

\bibliography{jss4560}

\appendix
\section[Implementation details]{Implementation details}

\subsection[Expression tree]{Expression tree}

A goal for \pkg{salmon} was to be able to represent variables in an object oriented fashion that also allowed easy manipulations / transformations. The solution to this was to represent the variables as an abstract syntax tree, borrowing heavily from a programming languages approach. Every expression of variables can be modeled as a tree where each inner node is an operation such as addition, multiplication, or transformation and each leaf is a variable (either quantitative or categorical). The following code and figure are equivalent in representation.

\begin{CodeInput}
>>>Q("a") - C("b") * C("c") + Q("d") ** 3 * Q("e") + \ 
   Sin(Q("f") ** 5) * Log(Q("g") + Q("h") + 1)
\end{CodeInput}

\begin{figure}[H]
\centering
\includegraphics{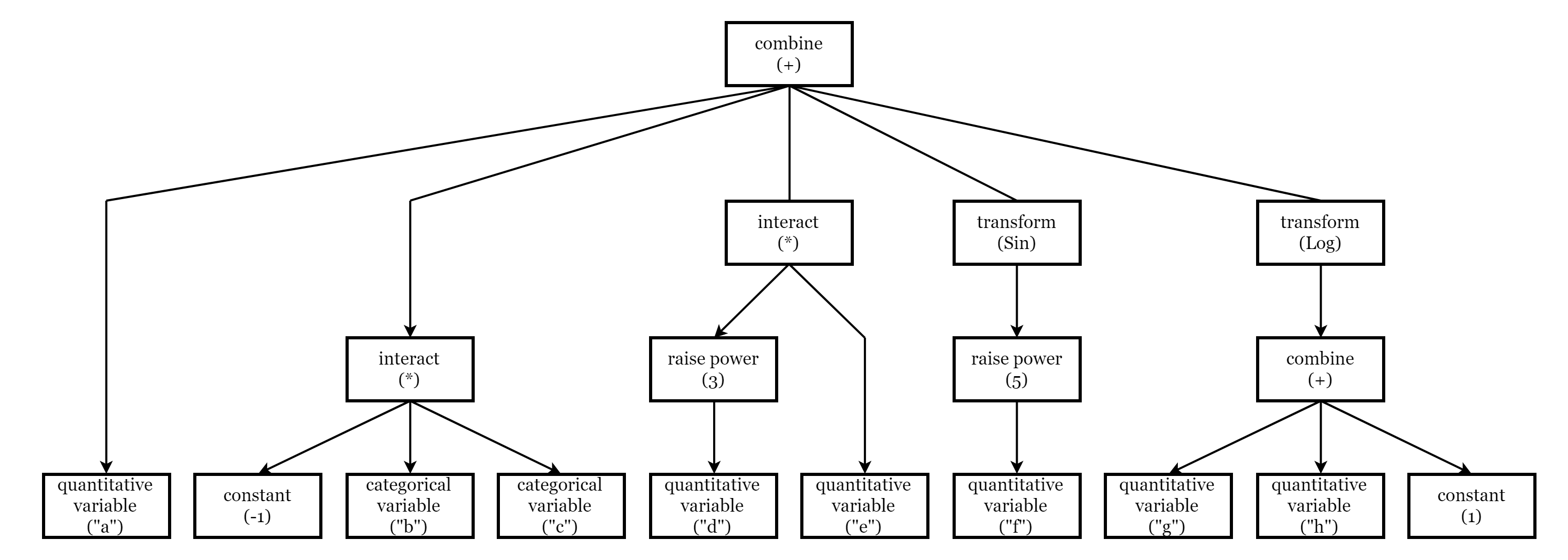}
\label{tree}
\end{figure}

This design allowed for a recursive style implementation when defining actions amongst variables. We denote the class \code{Expression} to denote the most general and all encompassing structure of variables. The exact object composition scheme can be described in BNF with the following grammar:

\begin{CodeInput}
<Expression>     ::= <Variable>
                   | <Constant>
                   | <Expression> + <Expression>
                   | <Expression> - <Expression>
                   | <Expression> * <Expression>
                   | <Expression> / <Expression>
                   | <Expression> ** <Constant>

<Variable>       ::= Var(<Name>)
                   | <Quantitative>
                   | <Categorical>
                   | <Transformation>(<Expression>)
                   | <Variable> ** <Constant>
                   | <Variable> * <Constant>
                   | <Variable> / <Constant>
                   | <Variable> * <Variable>
                   | <Variable> / <Variable>
                   
<Quantitative>   ::= Q(<Name>)
                  
<Categorical>    ::= C(<Name>)

<Transformation> ::= Sin
                   | Cos
                   | Exp
                   | Log
                   | Log10
                   | Standardize
                   | Center
                   | Identity

<Constant>       ::= int
                   | float
                   | <Constant> + <Constant>
                   | <Constant> - <Constant>
                   | <Constant> * <Constant>
                   | <Constant> / <Constant>

<Name>           ::= str
\end{CodeInput}

\subsection[Model fitting]{Model fitting}
\label{sec:model_fitting}

Our general strategy for fitting the linear regression model is similar to
\proglang{R}'s. \citep{R} We first calculate a QR decomposition 
of the design matrix $X$, which allows us to calculate the 
regression coefficients $\hat\bbeta$ 
by solving the system $R \hat\bbeta = Q^\top \by$ and the estimated covariance matrix 
$\hat\Sigma = \hat\sigma^2 (X^\top X)^{-1}$ by solving the system 
$R^\top R \hat\Sigma = I$ (since $X^\top X = R^\top R$).

When the model includes an intercept, \proglang{R} calculates the coefficients 
by appending a column of ones to the design matrix
\[ \tilde X = \begin{pmatrix} \ones & X \end{pmatrix}. \]
The regression coefficients are then given by 
\[ \hat\bbeta := \begin{pmatrix} \hat\beta_0 \\ \hat\bbeta_{1:p} \end{pmatrix} = (\tilde X^\top \tilde X)^{-1} \tilde X^\top \by. \]
 We take a slightly different approach, 
centering all of the variables first. The centered versions 
of the design matrix $X$ and the response variable $\by$ are given by 
\begin{align*}
X_c &= X - \frac{1}{n} \ones \ones^\top X & \by_c &= \by - \frac{1}{n} \ones \ones^\top \by \\
&:= X - \ones \bar\bx^\top & &:= \by - \ones \bar y
\end{align*}
It is well-known that centering does not change the coefficients of the 
explanatory variables. \citep{chatterjee2015regression} 
That is, we can calculate $\hat\beta_{1:p}$ as
\[ \hat\beta_{1:p} = (X_c^\top X_c)^{-1} X_c^\top \by_c \]
and recover the intercept as
\[ \hat\beta_0 = {\bf y} - \overline{{\bf x}}^\top \hat\beta_{1:p}. \]
The advantage of centering is that the computations are more 
numerically stable.

To calculate the covariance matrix, $\hat\Sigma$, the main hurdle is computing 
$(\tilde X^\top \tilde X)^{-1}$. We compute this without explicitly constructing 
$\tilde X$, using properties of block matrices.  To our knowledge, 
these formulas do not appear in the literature, 
so we reproduce the derivations here for completeness.

First, observe that the inverse of $\tilde X^\top \tilde X$ can be written in 
block form as
\begin{align*}
    (\tilde X^\top \tilde X)^{-1} &= \begin{pmatrix} n & \ones^\top X \\ X^\top \ones & X^\top X \end{pmatrix}^{-1} = \begin{pmatrix} a & {\bf b}^\top \\ {\bf b} & C \end{pmatrix}.
\end{align*}
Next, $C$ can be obtained as the inverse of the Schur complement of the block 
matrix $\tilde X^\top \tilde X$:
\begin{align*}
    C &= \Big(X^\top X - (X^\top \ones) \frac{1}{n} (\ones^\top X)\Big)^{-1} \\
    &= \Big(X^\top (I - P) X\Big)^{-1} & \text{($P := \frac{1}{n} \ones \ones^\top$ is a projection matrix)} \\
    &= \Big(X^\top (I - P)^2 X\Big)^{-1} & ((I-P)^2 = I - P) \\
    &= \Big(((I - P)X)^\top (I - P) X\Big)^{-1} & \text{($I - P$ is symmetric)} \\
    &= (X_c^\top X_c)^{-1}.
\end{align*}

Finally, we compute $a$ and ${\bf b}$ in terms of $C$.
\begin{align*}
    {\bf b} &= -C (X^\top \ones) \frac{1}{n} \\
    &= -C \bar\bx \\
    a &= \frac{1}{n} + \frac{1}{n} (\ones^\top X) C (X^\top \ones) \frac{1}{n} \\
    &= \frac{1}{n} - \bar \bx^\top {\bf b}.
\end{align*}

\end{document}